\documentclass[letterpaper, 10 pt, conference]{ieeeconf}  

\IEEEoverridecommandlockouts                              

\usepackage{amsmath} 
\usepackage{amssymb}  
\usepackage{caption}
\usepackage{subcaption}
\usepackage{graphicx}
\usepackage{algorithm}
\usepackage{algpseudocode}
\usepackage[utf8]{inputenc}
\usepackage{hyperref}
\usepackage{cite} 
\providecommand{\keyword}[1]
{
  \small	
  \textbf{\textit{Keywords---}} #1
}

\title{\LARGE \bf
Closed-Loop Control Law for Low Thrust Orbit Transfer with Guaranteed Stability }
\author{Suraj Kumar$^{1}$ $^{*}$, Aditya Rallapalli$^{1}$, Nivriti Priyadarshini$^{1}$, Bharat Kumar GVP$^{1}$, Ravi Kumar L$^{1}$
\thanks{$^*$Corresponding Author}
\thanks{$^{1}$The authors are associated with Controls and Digital Area, U R Rao Satellite Center, Indian Space Research Organization, Bengaluru, Karnataka, India \{surajk, adityar, bharat,nivriti, rkkumarl\}@ursc.gov.in}
}

\begin{document}

\maketitle
\thispagestyle{empty}
\pagestyle{empty}

\begin{abstract}

Electric propulsion is used to maximize payload capacity in communication satellites. These orbit raising maneuvers span several months and hundreds of revolutions, making trajectory design a complex challenge. The literature typically addresses this problem using feedback laws, with Q-law being one of the most prominent approaches. However, Q-law suffers from closed-loop stability issues, limiting its suitability for real-time on-board implementation. In this work, we focus on closed-loop orbit raising rather than offline trajectory planning and address the stability limitations of the Q-law through a Lyapunov based control design. A Lyapunov-guided modification of the classical Q-law is proposed to ensure closed-loop stability and enable real-time implementation. The effectiveness of the proposed method is demonstrated through closed-loop orbit transfers across various scenarios, including co-planar transfers, equatorial to polar orbit transfers, and geostationary transfer orbit (GTO) to geostationary earth orbit (GEO) transfers.
\end{abstract}
\keyword{\small Lyapunov Stability, Closed-loop Control, Orbit Transfer, Electric Propulsion}

\section{INTRODUCTION}
Electric propulsion systems are increasingly favored in communication satellites due to its high specific impulse—about ten times that of chemical thrusters—enabling greater payload capacity and large $\Delta{V}$ maneuvers. However, the resulting orbit transfer span hundreds to thousands of revolutions over several months, making trajectory design and execution highly complex, especially when accounting for eclipses, perturbations, and system uncertainties. Traditional open-loop planning and execution strategies require continuous ground intervention, leading to significant operational costs, extended deployment of manpower, and reliance on ground station support. In contrast, adopting a closed-loop orbit transfer strategy can greatly ease mission planning by reducing dependence on ground based corrections and enabling more autonomous spacecraft operations.

Trajectory design for low-thrust, multi-revolution transfer has been extensively studied for over two decades. Broadly, trajectory design approaches fall into two categories: (1) direct and indirect optimization methods, and (2) feedback control law based heuristics. Indirect methods apply optimal control theory and necessary conditions for optimality to solve minimum-time and/or minimum fuel problems\cite{kluever2010low, falck2012optimization}, while direct methods discretize the problem and solve it using nonlinear programming\cite{betts2000very,graham2016minimum, leomanni2021optimal}. Closed-loop feedback-driven (CLFD) control laws have gained popularity due to their low computational cost, making them suitable for rapid prototyping and onboard autonomous operations. These methods are typically based on either Lyapunov control theory or thrust blending strategies. Notable Lyapunov-based methods include Petropoulos’ Q-law\cite{petropoulos2004low, petropoulos2005refinements}, Joseph’s control law \cite{joseph2006lyapunov}, Naasz’s controllers \cite{naasz2002classical}, and Locoche law \cite{locoche2018analytical}, while Directional Adaptive Guidance \cite{falck2014comparison}  employs thrust blending strategy. Among these, Q-law\cite{petropoulos2005refinements} remains the most cited CLFD method. It has been extended in various ways—for example, Varga\cite{varga2016many} reformulated it in the equinoctial orbital frame to handle singularities, while others have combined Q-law with evolutionary algorithms \cite{lee2005low} for improved trajectory planning. Shannon \cite{shannon2020q} further integrated Q-law into a nonlinear programming framework to optimize gain selection and enforce nonlinear constraints on the initial state.

Q-law has been instrumental in planning low-thrust, multi-revolution trajectory transfers. However, the classical Q-law faces limitations for closed-loop onboard implementation due to its lack of formal stability guarantees. Although several heuristic modifications have been proposed in the literature to address these issues, they are typically validated only through simulations and lack theoretical rigor—sufficient for planning, but not for real-time control. This paper proposes a modified Q-law controller, a Lyapunov-based reformulation of the classical Q-law that ensures closed-loop stability and is suitable for real-time orbit control. 

The main contributions of this paper are the theoretical development of the controller, analysis of its stability properties, and demonstration of its effectiveness in closed-loop orbit transfer scenarios.

The paper is organized as follows: Section \ref{sec:bck} reviews the theoretical background, classical Q-law and its limitations. Section \ref{sec:mod-qlaw} presents the modified Q-law for closed-loop orbit transfer. Section \ref{sec:results} presents the simulation studies and comparison with existing literatures. Section \ref{sec:conc} concludes the paper.

\section{BACKGROUND}
\label{sec:bck}
\subsection{Dynamics}
The spacecraft's motion is described using Classical Orbital Elements (COE), and its transfer dynamics are governed by the Gauss variational equations:

\begin{equation}
\begin{aligned}
    \label{gauss_dyn}
    \frac{d\boldsymbol{Z}}{dt} = \begin{bmatrix}
        & \boldsymbol{\Phi}(\boldsymbol{Z}) & \\
        \frac{p\cos{\theta}}{eh} & -\frac{(p+r)\sin{\theta}}{eh} & 0
    \end{bmatrix}
    \boldsymbol{u} + 
    \begin{bmatrix}
        0 \\ \frac{h}{r^2}
    \end{bmatrix} \\
\end{aligned}   
\end{equation}
\begin{equation}
\label{stm}
    \begin{aligned}
    \boldsymbol{\Phi}(\boldsymbol{Z}) = \begin{bmatrix}
        \frac{2a^2}{h}e\sin{\theta} &  \frac{2a^2}{h}\frac{p}{r} & 0 \\
        \frac{p\sin{\theta}}{h} & \frac{(p+r)\cos{\theta}+re}{h} & 0 \\
        0 & 0 & \frac{r\cos(\theta+\omega)}{h} \\
        0 & 0 & \frac{r\sin(\theta+\omega)}{h\sin{i}} \\
        \frac{-p\cos{\theta}}{eh} & \frac{(p+r)\sin{\theta}}{eh} & \frac{r\sin(\theta+\omega)\cos{i}}{h\sin{i}} 
    \end{bmatrix}
    \end{aligned}
\end{equation}

where $\boldsymbol{Z} = [a, e, i, \Omega, \omega]^T$ is the vector of five orbital elements except true anomaly; $a$ is the semi-major axis; $e$ is the eccentricity; $i$ is the inclination; $\omega$ is the argument of perigee; $\Omega$ is the right ascension of ascending node (RAAN); $\theta$ is the true anomaly; $p$ is the semi-latus rectum; $h$ is the specific angular momentum; $r$ is the radius of spacecraft from central body; $\boldsymbol{u} \in \mathcal{R}^3$ denotes the perturbing acceleration vector in radial, tangential and normal direction respectively. True anomaly is treated separately from the other orbital elements, as it is typically considered a free parameter during the transfer. The perturbing acceleration comprises both the thrust generated by the onboard propulsion system and environmental disturbances such as Earth’s oblateness (J2 effect), solar radiation pressure, aerodynamic drag, and other perturbations. The onboard propulsion is assumed to produce constant thrust, as is typical for electric propulsion systems. We define $\dot{z}_{xx}$ as the maximum rate of change of the orbital element $z \in \boldsymbol{Z}$, taken over both the true anomaly of the osculating orbit and all possible thrust directions:
\begin{equation}
    \dot{z}_{xx} =
\begin{cases}
\max_{\alpha,\beta,\theta}\left( \dot{z} \right), & \text{for } z \in \{a, e, i,\Omega\} \\
\frac{\dot{\omega}_{xxi} + b\dot{\omega}_{xxo}}{1 + b}, & \text{for } z = \omega
\end{cases}
\end{equation}
where b is non-negative constant, $\alpha, \beta$ are the thrust vector angles represented in spherical coordinate and 
\begin{equation}
    \begin{aligned}
      \dot{\omega}_{xxi} &= \max_{\alpha,\theta}\left( \dot{\omega}|_{\beta=0} \right)   \\
      \dot{\omega}_{xxo} &= \max_{\theta}\left( \dot{\omega}|_{\beta=\frac{\pi}{2}} \right)
    \end{aligned}
\end{equation}
Detailed expressions for the maximum rate of change in each element can be computed analytically\cite{petropoulos2004low}. 
\subsection{Q-law Controller}
Q-law has been instrumental in planning low thrust multiple revolution trajectory design. Q-Law is typically used with an evolutionary algorithm to optimize control parameters to achieve more optimal solutions for planning time optimal transfer trajectory. The Q-law Lyapunov function is defined as
\begin{equation}
    Q = (1 + W_p P) \sum_{z \in \boldsymbol{Z}} W_z(\boldsymbol{Z}) S_z(\boldsymbol{Z}) \left(\frac{d(z,z_T)}{\dot{z}_{xx}}\right)^2 
\end{equation}
$\boldsymbol{Z_T} = [a_T, e_T, i_T, \Omega_T, \omega_T]^T$ is the vector of desired orbital elements; $\boldsymbol{W}$ is the non-negative gain vector associated with transfer; $d(z,z_T)$ is the distance between osculating and target element; $\boldsymbol{S}$ and scalar function $P$ are the scaling and penalty function that keeps semi-major axis and perigee height from growing too large or too small. These functions are defined as
\begin{equation}
\label{eqn:S}
\boldsymbol{S} =
\begin{cases}
\left( 1 + \left(\frac{a-a_T}{ma_T}\right)^n \right)^{\frac{1}{r}}  &  z = a \\
1 & z = e, i, \omega, \Omega
\end{cases}
\end{equation}
\begin{equation}
P = \exp\left(
\begin{bmatrix}
k \left(1 - \frac{r_p}{r_{p,\min}} \right)
\end{bmatrix}
\right)
\label{eq:P_definition}
\end{equation}
where $r_p$ is the periapsis radius; $r_{p,min}$ is the allowable minimum periapsis radius; $m,n,r,k$ are the hyperparameters with nominal values: $m=3, n=4, r=2, k=100$.

The Lyapunov function $Q$ is differentiated to minimize $\dot{Q}$, therefore driving $Q$ to zero as quickly as possible to the target. The derivative of Q is given as,
\begin{equation}
    \frac{dQ}{dt} = \sum_{z \in \boldsymbol{Z}} \frac{\partial Q}{\partial z} \frac{dz}{dt}
\end{equation}
Vectorially, it is expressed as
\begin{equation}
    \dot{Q} = \mathbf{Q}_z^T \Phi(\boldsymbol{Z})\boldsymbol{u}
\end{equation}
where $\mathbf{Q}_z \in \mathcal{R}^5$ is the column vector of partial derivatives of Q with respect to each orbital element. The thrust vector direction that minimizes $\dot{Q}$ is computed analytically:
\begin{equation}
\label{argminQdot}
    \boldsymbol{\hat{u}}^* = \arg \min_{\boldsymbol{\hat{u}}} \dot{Q} = -\frac{\boldsymbol{\Phi}(\boldsymbol{Z})^T \mathbf{Q}_z}{\left\| \boldsymbol{\Phi}(\boldsymbol{Z})^T \mathbf{Q}_z \right\|}
\end{equation}

\subsection{Limitations of Classical Q-law}
 The classical Q-law faces limitations for closed-loop onboard implementation. First, it requires computationally intensive numerical evaluation of derivatives. While symbolic computation of these derivatives is feasible for ground-based planning, onboard evaluation becomes challenging due to the complex algebraic and trigonometric expressions involved, particularly in terms like $\dot{z}_{xx}$. These expressions are computationally expensive and difficult to evaluate in real-time on low-speed onboard computers. This issue is especially relevant for electric propulsion system (EPS)-based missions such as GTO-to-GEO transfers, where onboard processors are typically slower and shared across multiple subsystems such as navigation, sensor processing, telemetry etc. In such scenarios, evaluating the required derivatives within the available control cycle time becomes impractical.

The classical Q-law also suffers from stability issues. Various heuristic modifications have been introduced to address these, but they lack formal theoretical guarantees and are typically validated only through simulations. For example, without the inclusion of the scaling function $S$ for the semi-major axis term, the Lyapunov function used in the Q-law framework does not satisfy the Lyapunov stability criterion. Although the form of $S$ as shown in Eq (\ref{eqn:S}) has been shown to work well in many simulations, the absence of a rigorous theoretical foundation remains a concern. Specifically, there is no guarantee that the control law will not direct the spacecraft toward an unbounded increase in semi-major axis instead of converging toward the target orbit. Another issue arises from the complexity of the partial derivatives of the Q-function with respect to the orbital elements, which can lead to unintended control behavior that is not immediately evident, as discussed in detail by Hatten in \cite{hatten2012critical}. Notably, the derivatives of the semi-major axis and inclination terms in Q with respect to eccentricity are always negative, implying that increasing eccentricity will always reduce the derivative of Lyapunov function with respect to semi-major axis and inclination. As a result, a maneuver targeting only the semi-major axis may not produce acceleration strictly in the direction of velocity, and a maneuver targeting only the inclination may not yield purely out-of-plane acceleration, contrary to intuitive expectations. In fact, the controller often demands large changes in eccentricity for both semi-major axis and inclination transfers. This tendency is one of the main reasons why Petropoulos introduced a minimum-periapsis-distance penalty function $P$ into the control law to avoid collision with central body. Another critical limitation of the Q-law is its lack of control over the sacrificial overshoot in the semi-major axis during inclination change maneuvers. When coast arcs are not permitted, Q-law induces a significant increase in the semi-major axis, which can reduce fuel consumption by enabling more efficient inclination changes at higher altitudes. However, this overshoot is not inherently bounded by the control law, and in certain scenarios—even with the inclusion of the scaling function—this can lead to an unbounded growth in the semi-major axis \cite{hatten2012critical}.
\section{Modified Q-law Controller}
\label{sec:mod-qlaw}
The Q-function can be viewed as a weighted sum of optimistic transfer times required to reach the target orbital elements. For each element \( z \in \boldsymbol{Z} \), the term
$\frac{d(z,z_T)}{\dot{z}_{xx}}$ represents the minimum time needed to drive the element from its current value to the target \( z_T \), where \( d(z, z_T) \) denotes the absolute distance in the element space, and \( \dot{z}_{xx} \) is the maximum rate in each orbital element. This rate represents the best-case local velocity, and thus the ratio represents the minimum transfer time required for each orbital element. A key advantage of such a function is that the each element in the Q function will decrease not only if \( d(z, z_T) \) decreases but also if \( \dot{z}_{xx} \) increases. This allows the control law to perform sacrificial correction in orbital elements - for instance, the optimal rate of change of inclination increases in magnitude as semi-major axis increases. However, a notable limitation of this structure lack of closed-loop Lyapunov stability and the sacrificial corrections in orbital elements are unbounded by design. Therefore, we propose modifications for candidate Lyapunov function  associated with semi-major axis, eccentricity and inclination transfer. The argument of perigee and RAAN is not considered in the present study. Nonetheless, the approach described here can be extended to include corrections in the argument of perigee transfer and RAAN as needed. The proposed candidate Lyapunov term for each orbital element is given as
\begin{equation}
\tilde{V}_z = \tilde{K}_z (z - z_T)^2, z \in {a,e,i} 
\end{equation}
where where \( \tilde{K}_z \) is a state-dependent term associated with the orbital element transfer, defined as:
\begin{equation}
\tilde{K}_z = \frac{1}{{\tilde{\dot{z}}_{xx}}^2}
\end{equation}
The symbol $\tilde{x}$ denotes appropriate modification in the generic function or variable $x$. This structure has the advantage that the original and modified control law represented in unified notation. The modified Lyapunov function for the complete transfer is then composed of weighted combination of the individual Lyapunov term:
\begin{equation}
    \tilde{V} = \sum_{z \in \boldsymbol{Z}} W_z\tilde{K}_z(\boldsymbol{Z}) d(z,z_T)^2
\end{equation}
The Lyapunov function $\tilde{V}$ is differentiated to minimize $\dot{\tilde{V}}$, therefore driving $\tilde{V}$ to zero as quickly as possible to the target. The derivative of $\tilde{V}$ is given as,
\begin{equation}
    \frac{d\tilde{V}}{dt} = \sum_{z \in \boldsymbol{Z}} \frac{\partial \tilde{V}}{\partial z} \frac{dz}{dt}
\end{equation}
Vectorially, it is expressed as
\begin{equation}
    \dot{\tilde{V}} = \tilde{\mathbf{V}}_z^T \Phi(\boldsymbol{Z})\boldsymbol{u}
\end{equation}
where $\tilde{\mathbf{V}}_z \in \mathcal{R}^5$ is the column vector of partial derivatives of $\tilde{V}$ with respect to each orbital element. The thrust vector direction that minimizes $\dot{\tilde{V}}$ is computed analytically:
\begin{equation}
\label{argminQdot}
    \boldsymbol{\hat{u}}^* = \arg \min_{\boldsymbol{\hat{u}}} \dot{\tilde{V}} = -\frac{\boldsymbol{\Phi}(\boldsymbol{Z})^T \tilde{\mathbf{V}}_z}{\left\| \boldsymbol{\Phi}(\boldsymbol{Z})^T \tilde{\mathbf{V}}_z \right\|}
\end{equation}
We now present the modifications for each orbital element.
\subsection{Semi-major Axis Transfer:}
Consider the Lyapunov term for semi-major axis transfer 
\begin{equation}
V_a = \left(\frac{a-a_T}{\dot{a}_{xx}}\right)^2
\end{equation}
where $f$ denotes the magnitude of acceleration and $\dot{a}_{xx} = 2 f \sqrt{\frac{a^3 (1+e)}{\mu (1-e)}}$ 
On expanding and taking derivative of $V_a$ w.r.t $a$ and $e$, we get
\begin{equation}
\label{eq:orig_Vaderiv}
\begin{aligned}
    V_a & = c \cdot \frac{1}{4a^3} \cdot \frac{1 - e}{1 + e} \cdot (a - a_T)^2 \\
    \frac{\partial V_a}{\partial a} &= \frac{c}{4a^4} \cdot \frac{1 - e}{1 + e} \cdot (a - a_T)(-a + 3a_T) \\
    \frac{\partial V_a}{\partial e} &= - \frac{c}{4a^3} \cdot \frac{2}{(1 + e)^2} \cdot (a - a_T)^2 \leq 0
\end{aligned}
\end{equation}
where $c = \frac{\mu}{f^2}$. The Lyapunov term \( V_a \) reaches zero not only at the desired target \( a = a_T \), but also asymptotically as \( a \to \infty \), indicating that convergence to \( a_T \) is not inherently guaranteed. Additionally, the partial derivative \( \frac{\partial V}{\partial a} \) vanishes at \( a = a_T \), \( a = 3a_T \), and as \( a \to \infty \), introducing multiple equilibrium points. While the control law derived from minimizing \( \frac{dV_a}{dt} \) ensures that \( V_a \) is non-increasing, this alone does not imply asymptotic convergence to \( a_T \). According to LaSalle’s invariance principle, convergence requires that the only invariant set where \( \frac{dV}{dt} = 0 \) corresponds to is the target state. However, the existence of a spurious equilibrium at \( a = 3a_T \) violates this condition and opens the possibility of undesired convergence. Moreover, the partial derivative of \( V_a \) with respect to eccentricity is always non-positive, meaning that increasing eccentricity always reduces \( V_a \). As a result, maneuvers aimed solely at correcting the semi-major axis may inadvertently cause unbounded growth in eccentricity. This unconstrained increase could violate minimum periapsis altitude constraints, potentially resulting in impact with the central body.

We now present the modifications to the Lyapunov term associated with semi-major axis transfer by appropriately modifying \( \tilde{\dot{a}}_{xx} \), approximation to the true best-case rate of change \( \dot{a}_{xx} \), as:
\begin{equation}
\begin{aligned}
\tilde{\dot{a}}_{xx} = 2f \Bigg[ &
\mathcal{I}(a < a^*) 
\sqrt{\dfrac{a^3(1+e)}{\mu(1-e)}} + \mathcal{I}(a \geq a^*) 
\sqrt{\dfrac{a^{*3}(1+e)}{\mu(1-e)}}  \Bigg] \\
\end{aligned}
\end{equation}
where $a^* = \zeta a_T, \quad \zeta \in (1, 3)$; \( \mathcal{I} \) is an indicator function that evaluates to 1 if the argument is true, and 0 otherwise. Beyond a prescribed threshold (set to be less than \( 3a_T \)), the dependency on the semi-major axis is intentionally removed. 
The parameter \( \zeta > 1 \) determines the threshold \( a^* \) beyond which this decoupling occurs. Overshoot in the semi-major axis often arises due to sacrificial corrections in other orbital elements (e.g., inclination). The value of \( \zeta \) limits the maximum allowable overshoot.  Beyond \( \zeta a_T \), the constant value of \( \tilde{\dot{a}}_{xx} \) (with respect to \( a \)) implies that the transfer proceeds at a constant rate, rather than increasing indefinitely with \( a \) as in the classical Q-law formulation.  With the proposed modification, the associated derivatives are given as:
\begin{equation}
\begin{aligned}
\dfrac{\partial \tilde{K}_a}{\partial a} &= -\frac{3\tilde{K}_a}{a} \, \mathcal{I}(a < a^*) \\
\dfrac{\partial \tilde{K}_a}{\partial e} &= 
    \frac{c}{4a^3} \left( -\frac{2}{(1+e)^2} \right) \mathcal{I}(a < a^*) \\
    &\quad + \frac{c}{4a^{*3}} \left( -\frac{2}{(1+e)^2} \right) \mathcal{I}(a \geq a^*) \\
\frac{\partial \tilde{V}_a}{\partial a} &= 
    2\tilde{K}_a(a - a_T) + (a - a_T)^2 \frac{\partial \tilde{K}_a}{\partial a} \\
\frac{\partial \tilde{V}_a}{\partial e} &= 
    (a - a_T)^2 \frac{\partial \tilde{K}_a}{\partial e} \, \mathcal{I}(r_p > r_{p,\min})
\end{aligned}
\end{equation}
The dependency of \( V_a \) on eccentricity is removed when the periapsis radius falls below a minimum acceptable value. This prevents sacrificial increases in eccentricity that could otherwise violate the minimum periapsis constraints. As a result, the thrust vector is reoriented purely along the velocity direction, avoiding any increase in eccentricity.

Partial derivatives with respect to other orbital elements is zero as $\tilde{K}_a$ only depends on semi-major axis and eccentricity. 
\subsection{Eccentricity Transfer:}
Consider the Lyapunov term for eccentricity transfer
\begin{equation}
V_e = \left( \frac{e-e_T}{\dot{e}_{xx}}\right)^2
\end{equation}
On expanding and taking derivative of $V_e$ w.r.t $a$ and $e$, we get
\begin{equation}
\begin{aligned}
    V_e &= c \cdot \dfrac{(e - e_T)^2}{4a(1 - e^2)} \\
    \dfrac{\partial V_e}{\partial a} &= -c \cdot \dfrac{(e - e_T)^2}{4a^2(1 - e^2)} \leq 0 \\
    \dfrac{\partial V_e}{\partial e} &= \dfrac{c}{2a} \cdot \dfrac{(e - e_T)\left(1 - 2e^2 + e e_T\right)}{(1 - e^2)^2}
\end{aligned}
\end{equation}
The Lyapunov function \( V_e \) reaches zero not only at the target eccentricity \( e = e_T \), but also as the semi-major axis \( a \to \infty \). Since \( \frac{\partial V_e}{\partial a} \) is always non-positive, increasing \( a \) continually reduces \( V_e \). However, \( \frac{\partial V_e}{\partial a} \) and \( \frac{\partial V_e}{\partial e} \) vanish only at \( e = e_T \), making it the sole equilibrium under LaSalle’s invariance principle. To eliminate the spurious minimum at infinity and prevent unbounded growth in \( a \) as a means to reduce \( e \), we modify the Lyapunov function as follows.

Consider the modified Lyapunov function for eccentricity transfer as
\begin{equation}
\tilde{V}_e = \tilde{K}_e (e - e_T)^2
\end{equation}
The proposed modification in \( \tilde{\dot{e}}_{xx} \), an approximation to the true best-case rate of change \( \dot{e}_{xx} \) is given as,
\begin{equation}
\label{Ke}
\begin{aligned}
    K_e &=
    \begin{cases} 
        \dfrac{c}{4\min(a,a_T)(1 - e^2)}, & \text{if } e \leq 1 - \delta_e \\
        \dfrac{c}{4\min(a,a_T)(1 - (1-\delta_e)^2)}, & \text{if } e > 1-\delta_e \\
    \end{cases} \\
\end{aligned}
\end{equation}
To avoid singularity in \( K_e \) as eccentricity approaches unity, a small threshold \( \delta_e \) is introduced. For \( e > 1 - \delta_e \), the sharp rise in \( K_e \) is replaced with a constant value. This regularization is crucial, as the Q-law inherently couples control inputs through derivatives like \( \partial V_a / \partial e \), which remains negative throughout, causing the original Q-function to drive \( e \) upward aggressively. Additionally, \( \partial V_e / \partial a \) remains non-positive, so increasing \( a \) consistently reduces \( V_e \). However, simulations reveal that this can cause an unbounded rise in \( a \), far exceeding the target \( a_T \), without effectively minimizing \( V_e \). Since \( V_e = 0 \) also holds as \( a \to \infty \), this introduces an undesirable equilibrium. To mitigate this, we set \( \partial V_e / \partial a = 0 \) once \( a \) exceeds \( \zeta a_T \), allowing eccentricity correction at a constant rate beyond the threshold. The factor \( \zeta \in (1, 3) \) thus bounds the sacrificial increase in \( a \) during eccentricity correction.The associated derivatives are given as

\begin{equation}
    \begin{aligned}
    \dfrac{\partial \tilde{K}_e}{\partial e} &= \dfrac{2\tilde{K}_e e}{1 - e^2}\mathcal{I}(e \leq 1 - \delta_e) \\
    \dfrac{\partial \tilde{K}_e}{\partial a} &= \dfrac{-\tilde{K}_e}{a} \mathcal{I}(a < a^*) \\
      \dfrac{\partial \tilde{V}_e}{\partial a} &= (e-e_T)^2 \dfrac{\partial \tilde{K}_e}{\partial a} \\
    \dfrac{\partial \tilde{V}_e}{\partial e} &= (e-e_T)^2 \dfrac{\partial \tilde{K}_e}{\partial a} + 2 \tilde{K}_e (e-e_T)
    \end{aligned}
\end{equation}
Partial derivatives with respect to other orbital elements is zero as $\tilde{V}_e$ only depends on semi-major axis and eccentricity.
\subsection{Inclination Transfer:}
Consider the Lyapunov term for inclination transfer
\begin{equation}
\begin{aligned}
V_i &= \left( \frac{i-i_T}{\dot{i}_{xx}}\right)^2
\end{aligned}
\end{equation}
On expanding we get,
\begin{equation}
\begin{aligned}
V_i &=  \frac{c}{a(1-e)^2} \mathcal{F}_i^2 (i-i_T)^2 \\  \mathcal{F}_i &= \left( \sqrt{1 - e^2 \sin^2(\omega)} - e \left| \cos(\omega) \right| \right)
\end{aligned}
\end{equation}
The Lyapunov function attains zero not only at the desired target value $i = i_T$ but also in the limit as $a \to \infty, e \to 1$ which is undesirable. 
We consider first order approximation for the term $ \sqrt{1 - e^2 \sin^2(\omega)}$ in the function $\mathcal{F}_i$:
\begin{equation}
    \begin{aligned}
        \tilde{\mathcal{F}}_i &= \left(1 - \frac{1}{2}e^2 \sin^2(\omega)) - e \left| \cos(\omega) \right| \right) \\
    \end{aligned}
\end{equation}
Following relation holds for the true function and its approximation:
\begin{equation}
    \begin{aligned}
        \tilde{\dot{i}}_{xx} \leq \dot{i}_{xx} \\
        \implies \tilde{V}_i \geq V_i 
    \end{aligned}
\end{equation}
As discussed earlier, since $\dot{i}_{xx}$ is only the best-case maximum rate of change in inclination, $ \tilde{\dot{i}}_{xx}$ is uniformly an under-approximation of the true function. Therefore, true function is replaced by this approximation. The function $\tilde{K}_i$ is given as,
\begin{equation}
\begin{aligned}
\tilde{K}_i &= 
\begin{cases}
\dfrac{c}{\min(a,a^*)(1 - e^2)} \left( \tilde{\mathcal{F}_i} \right)^2, & \text{if } e < 1 - \delta_e \\
\int_{1-\delta_e}^1 \left( \dfrac{c}{a(1 - e^2)} \left( \tilde{\mathcal{F}_i} \right)^2\right) de, & \text{otherwise}
\end{cases}
\end{aligned}
\end{equation}
$\delta_e$ is introduced similar to eccentricity term to avoid singularity as eccentricity approaches 1. Beyond a prescribed threshold, the dependency of $\tilde{K}_i$ on semi-major axis is intentionally removed to upper bound the sacrificial overshoot in semi-major axis for inclination correction. The value of $1 < \zeta < 3$ limits the allowable overshoot. Even though $\tilde{K}_i$ term depends on the eccentricity, sensitivity analysis showed inclusion of derivative of $\tilde{K}_i$ with respect to eccentricity is negligible. Therefore this term is forced to zero to avoid complex derivative computation. Partial derivatives of $\tilde{K}_i$ with respect to other orbital elements are given as,
\begin{equation}
\begin{aligned}
   \frac{\partial \tilde{K}_i}{\partial a} &= -\frac{\tilde{K}_i}{a}\mathcal{I}(a<a^*) \\
   \frac{\partial \tilde{V}_i}{\partial a} &= (i - i_T)^2 \cdot \frac{\partial \tilde{K}_i}{\partial a} \\
    \frac{\partial \tilde{V}_i}{\partial e} &= 0 \quad \text{(derivative forced to 0)} \\
    \frac{\partial \tilde{V}_i}{\partial i} &= 2(i - i_T)\tilde{K}_i \\
     \frac{\partial \tilde{K}_i}{\partial \omega} &= \frac{\mu}{f^2} \cdot \frac{1}{a(1 - e^2)} \cdot 2\tilde{\mathcal{F}}_i \cdot \frac{\partial \tilde{\mathcal{F}}_i}{\partial \omega} \\
     \frac{\partial \tilde{V}_i}{\partial \omega} &= (i - i_T)^2 \cdot \frac{\partial \tilde{K}_i}{\partial \omega}
\end{aligned}
\end{equation}

\subsection{Relative Effectivity and Coasting Arcs}

The formulation described above provides a framework suitable for solving near minimum-time low-thrust transfer problems, as thrusting is applied continuously throughout the orbit. However, to achieve minimum-propellant transfers at the expense of increased transfer duration, the classical Q-law introduces a coasting mechanism that switches off the engines at locations along the orbit where thrusting is inefficient. This behavior is governed by the relative effectivity, a scalar quantity that indicates how effective the thrust is at the current true anomaly compared to other locations on the osculating orbit, and is defined as

\begin{equation}
\eta_r = \frac{\dot{\tilde{V}}_n - \dot{\tilde{V}}_{nx}}{\dot{\tilde{V}}_{nn} - \dot{\tilde{V}}_{nx}}
\label{eq:eta_r}
\end{equation}

\begin{equation}
\dot{\tilde{V}}_{nn} = \min(\dot{\tilde{V}}_n)
\label{eq:Qnn}
\end{equation}

\begin{equation}
\dot{\tilde{V}}_{nx} = \max(\dot{\tilde{V}}n)
\label{eq:Qnx}
\end{equation}
Here, $\dot{\tilde{V}}_n, \dot{\tilde{V}}_{nn}, \dot{\tilde{V}}_{nx}$ denotes the current rate of change, minimum and maximum rate of change in lyapunov derivative in the osculating orbit respectively. Naturally, $\eta_r \in (0,1)$.  

To introduce coasting, non-zero value of $\eta_{r,thrshld}$ threshold is selected and thrusting is disabled as long as $\eta_r < \eta_{r,thrshld}$
 
\section{Results}  
\label{sec:results}
Low-thrust time-optimal transfer is simulated using a high-fidelity in-house simulation package that models Keplerian dynamics along with key environmental perturbations, including J2 effects, third-body perturbations, and solar radiation pressure. Sun model and eclipse arcs are also incorporated, during which the spacecraft coasts without thrust. Q-law, whether in its classical or modified form, can be used with arbitrary unit gain. However, it is well established that the performance of Q-law can be significantly improved by appropriately tuning the gain weights. In this work, Particle Swarm Optimization (PSO) is employed to tune the weights for both the classical and modified Q-law formulations, with the objective of minimizing transfer time for given relative effectivity threshold to generate Pareto-optimal curve showing trade-off between transfer time and propellant consumption. Three orbit transfer scenarios are considered to demonstrate the efficacy of the proposed approach and compare its performance with the classical Q-law. Tables \ref{tab:dyn_data} and \ref{tab:orbit-cases} list the dynamics simulation parameters for the different cases, as well as the initial and terminal orbits used in the simulations.
\begin{table}[hbt!]
\caption{\label{tab:dyn_data} Dynamics Simulation Parameters for Different Cases}
\centering
\begin{tabular}{lccc}
\hline
\textbf{Parameter} & \textbf{Case A} & \textbf{Case B} & \textbf{Case C} \\
\hline
Mass, kg              & 300   & 300    & 1200  \\
Thrust, N             & 1.0  & 1.0  & 0.312 \\
$I_{sp}$, s           & 3100   & 3100   & 1800  \\
\hline
\end{tabular}
\end{table}
\begin{table}[hbt!]
\caption{\label{tab:orbit-cases} Orbit Transfer Cases: Orbital Elements}
\centering
\begin{tabular}{lcccccc}
\hline
Case & Orbit & a(km) & e & i($\deg$) & $\Omega(\deg)$ & $\omega(\deg)$  \\
\hline
A & Initial & 7000 & 0.01 & 0.05 & 0.0 & 0.0 \\ 
  & Target  & 42000 & 0.01 & free & free & free \\ 
\hline
B & Initial & 10000 & 0.005 & 0.05 & 0.0 & 0.0 \\ 
  & Target  & 10000 & 0.005 & 90.0 & free & free \\ 
\hline
C & Initial & 24363.9 & 0.73 & 28.5 & 0.0 & 178.0 \\ 
  & Target  & 42164 & 0.01 & 0.01 & free & free \\ 
\hline
\end{tabular}
\end{table}
\subsection{Coplanar LEO to GEO Transfer}
\begin{figure}[hbt!]
  \centering
  \begin{minipage}{.23\textwidth}
    \centering
    \includegraphics[width=0.9\textwidth]{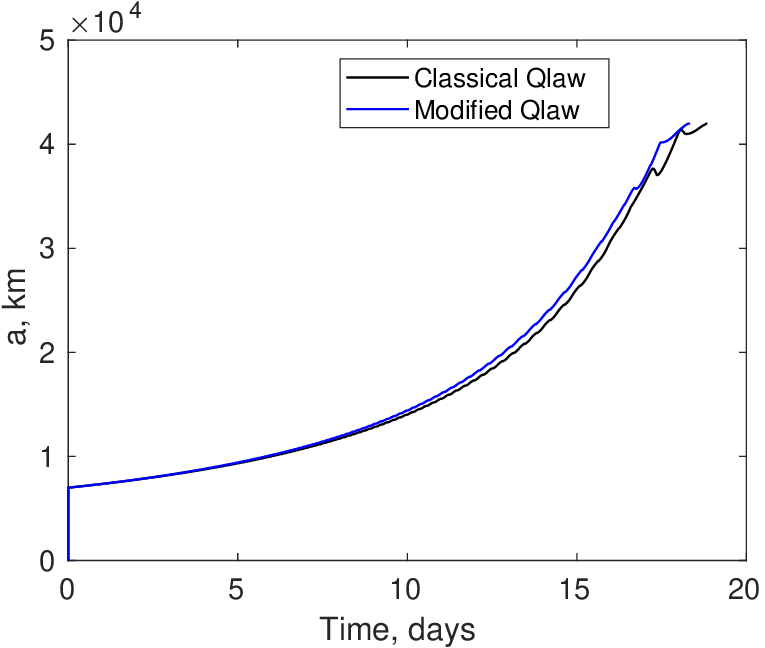}
    \subcaption{Semi-major axis}\label{fig:a_A}
  \end{minipage}
  \begin{minipage}{.23\textwidth}
    \centering
    \includegraphics[width=0.9\textwidth]{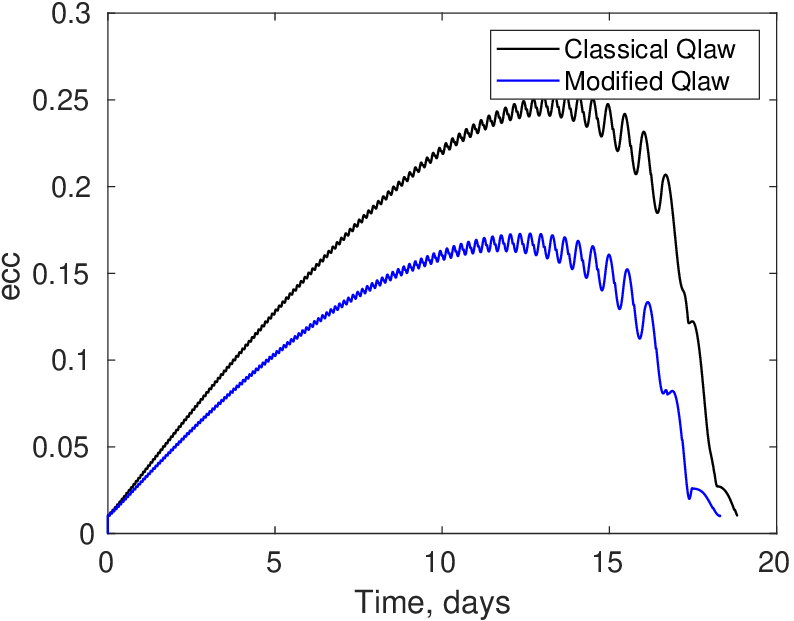}
    \subcaption{Eccentricity}\label{fig:ecc_A}
  \end{minipage}
  \begin{minipage}{.23\textwidth}
    \centering
    \includegraphics[width=0.9\textwidth]{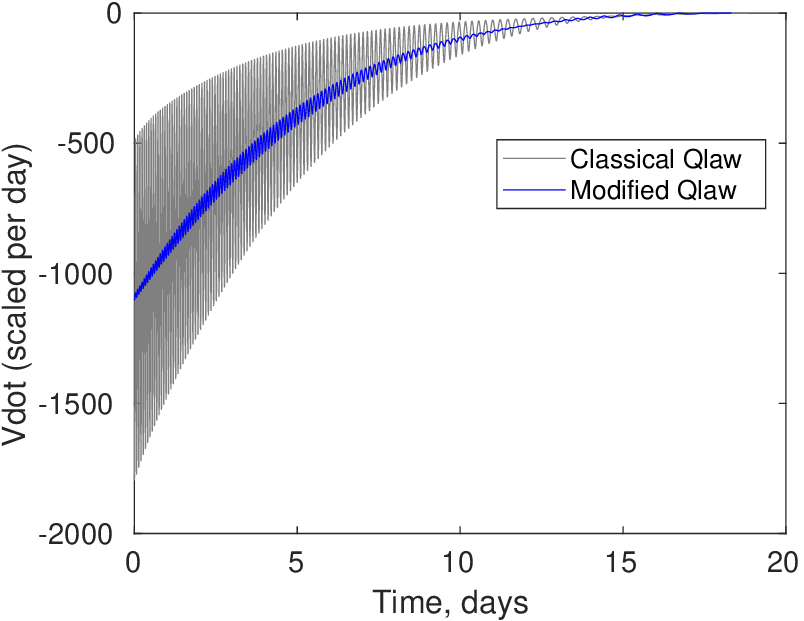}
    \subcaption{$\dot{V}$}\label{fig:vdot_A}
  \end{minipage}
    \begin{minipage}{.23\textwidth}
    \centering
    \includegraphics[width=0.9\textwidth]{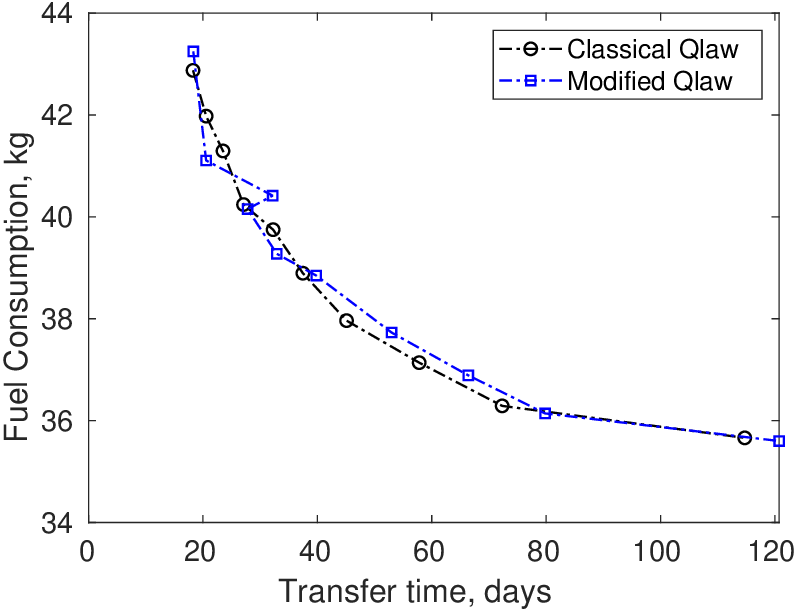}
    \subcaption{Pareto Front}\label{fig:pareto_front_A}
  \end{minipage}
  \caption{Coplanar Transfer:: Comparison between Classical and Modified Qlaw}
  \label{fig:coplanar}
\end{figure}

The in-plane circular-to-circular orbit raising problem is a well-studied case. Fig \ref{fig:a_A}–\ref{fig:a_B} show the comparison of the semi-major axis and eccentricity trajectories for both the classical Q-law and the modified Q-law, considering a relative effectivity threshold of 0.0, which corresponds to continuous thrusting and approximates a minimum-time solution. Fig \ref{fig:vdot_A} illustrates the evolution of the Lyapunov derivative for both control laws, while Fig \ref{fig:pareto_front_A} presents the Pareto front of transfer time versus propellant consumption. It can be observed that the modified Q-law achieves performance comparable to the classical Q-law. 

\subsection{Equatorial Orbit to Polar Orbit Transfer}
\begin{figure}[hbt!]
  \centering
  \begin{minipage}{.23\textwidth}
    \centering
    \includegraphics[width=0.9\textwidth]{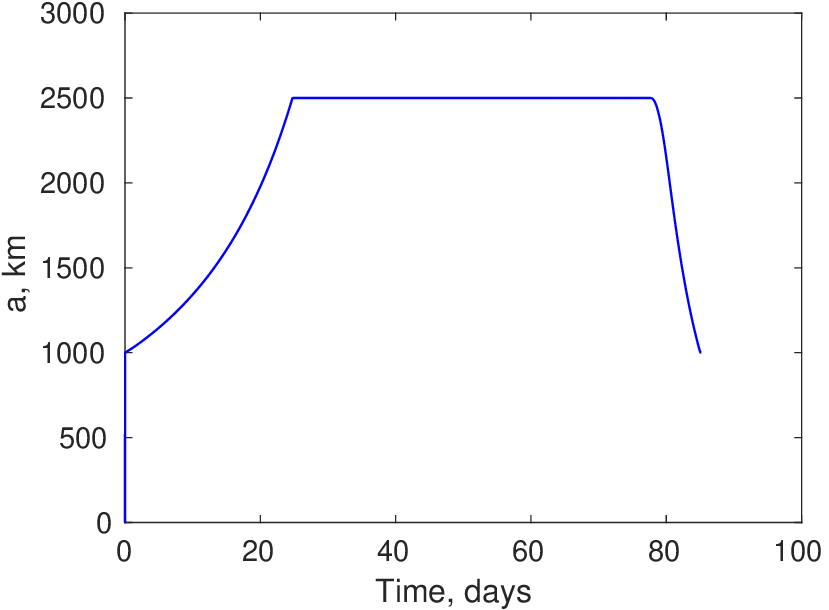}
    \subcaption{Semi-major axis}\label{fig:a_B}
  \end{minipage}
  \begin{minipage}{.23\textwidth}
    \centering
    \includegraphics[width=0.9\textwidth]{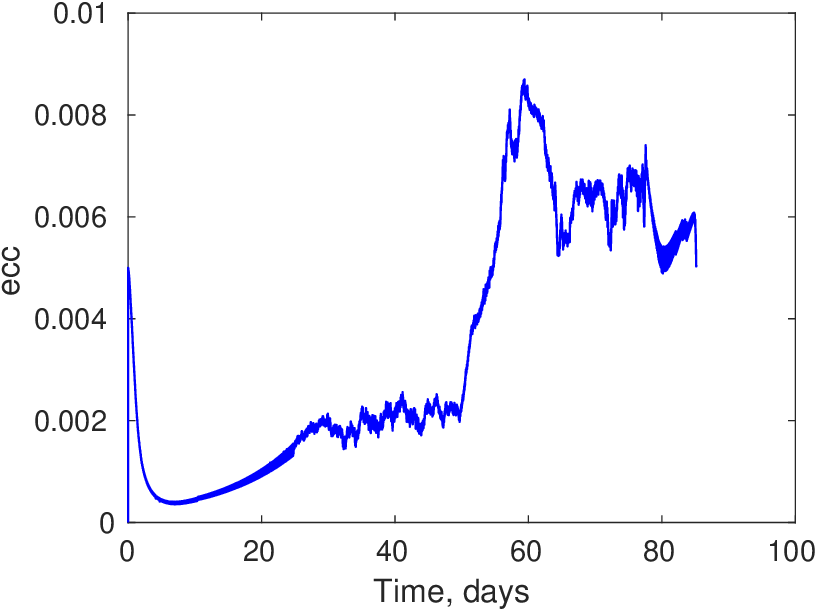}
    \subcaption{Eccentricity}\label{fig:ecc_B}
  \end{minipage}
  \begin{minipage}{.23\textwidth}
    \centering
    \includegraphics[width=0.9\textwidth]{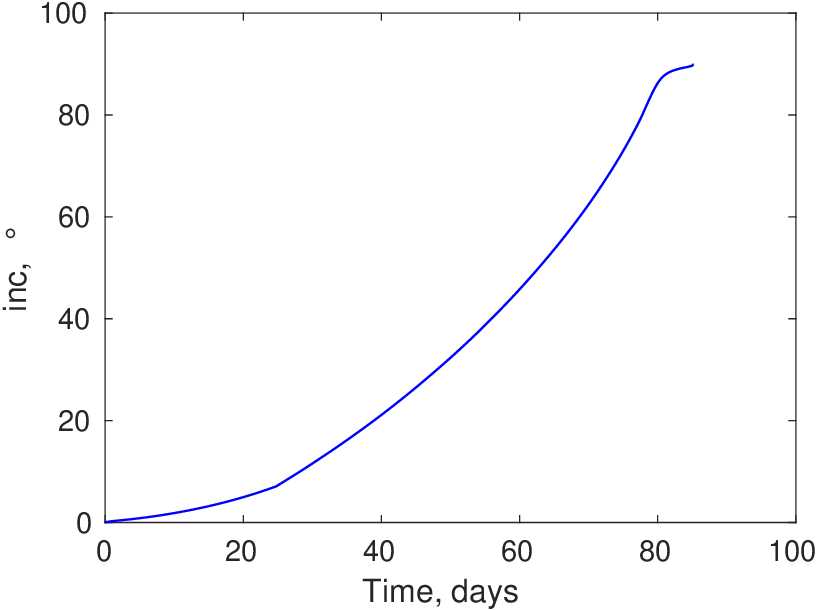}
    \subcaption{Inclination}\label{fig:inc_B}
  \end{minipage}
    \begin{minipage}{.23\textwidth}
    \centering
    \includegraphics[width=0.9\textwidth]{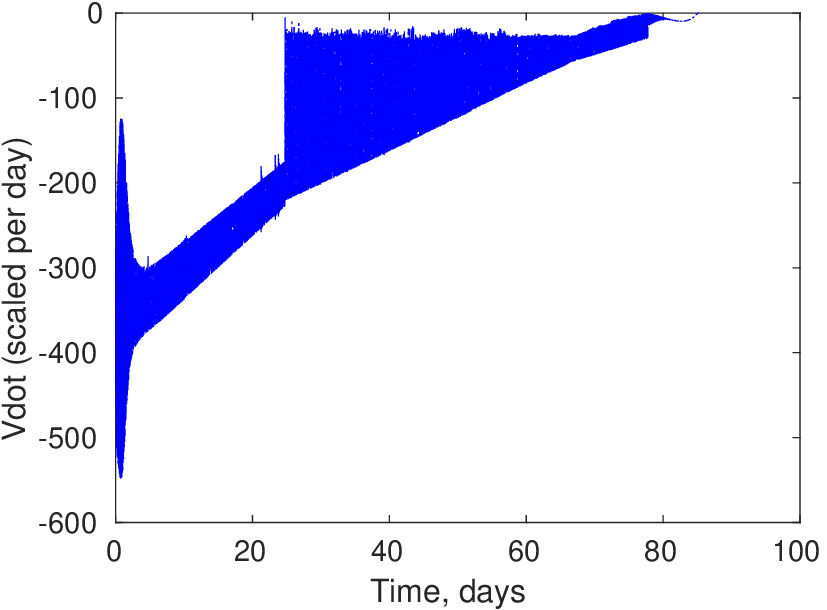}
    \subcaption{$\dot{\tilde{V}}$ Evolution}\label{fig:vdot_B}
  \end{minipage}
  \caption{Equatorial Orbit to Polar Orbit Transfer (Plots show trajectory trace for Modified Qlaw while classical Qlaw didn't converge for this scenario)}
  \label{fig:coplanar}
\end{figure}

This scenario presents an interesting case in which the classical Q-law fails to converge to the desired target orbit. Here, an inclination change of $90^\circ$ is required. As discussed earlier, performing inclination changes is more efficient at larger semi-major axes; however, the classical Q-law demands an unbounded increase in both the semi-major axis and eccentricity. As these values grow, the perigee radius decreases, ultimately causing the trajectory to collide with the central body, even when a penalty function is employed. The modified Q-law, on the other hand, effectively upper bounds this sacrificial correction by selectively removing the dependency on the semi-major axis, as illustrated in Figs. \ref{fig:a_B}–\ref{fig:inc_B}. Consequently, the transfer proceeds at a steady rate, ensuring proper convergence. Figure \ref{fig:vdot_B} shows the Lyapunov derivative, which remains strictly non-positive, thereby guaranteeing stability.

\subsection{GTO to GEO Transfer }
\begin{figure}[hbt!]
  \centering
  \begin{minipage}{.23\textwidth}
    \centering
    \includegraphics[width=0.9\textwidth]{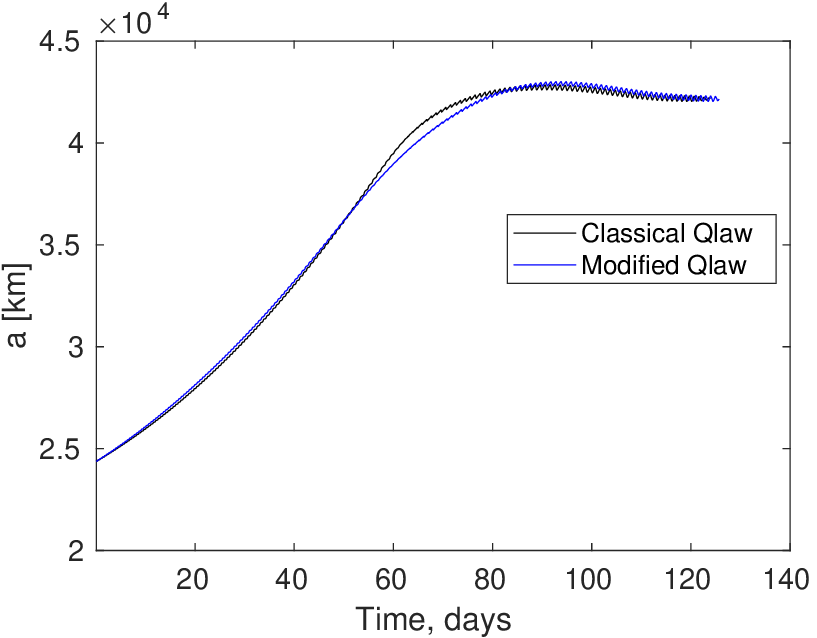}
    \subcaption{Semi-major axis}\label{fig:a_C}
  \end{minipage}
  \begin{minipage}{.23\textwidth}
    \centering
    \includegraphics[width=0.9\textwidth]{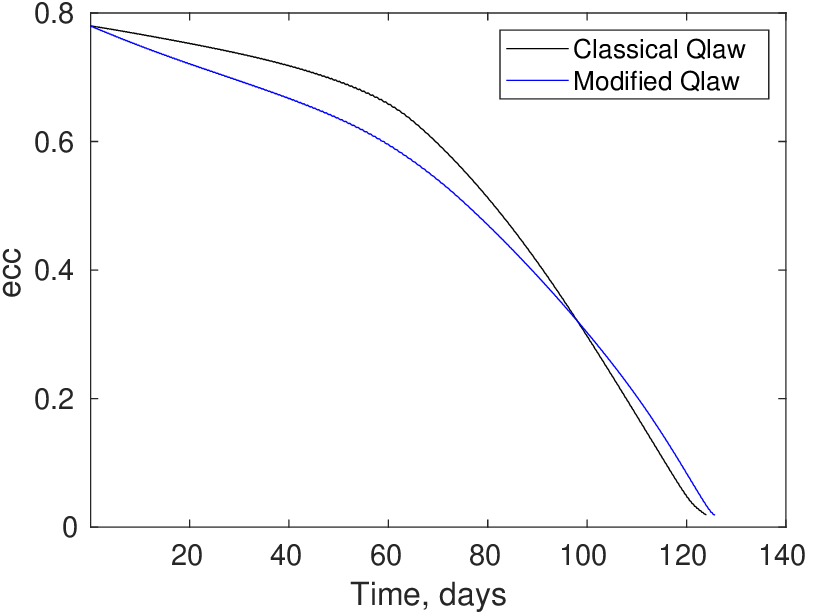}
    \subcaption{Eccentricity}\label{fig:ecc_C}
  \end{minipage}
  \begin{minipage}{.23\textwidth}
    \centering
    \includegraphics[width=0.9\textwidth]{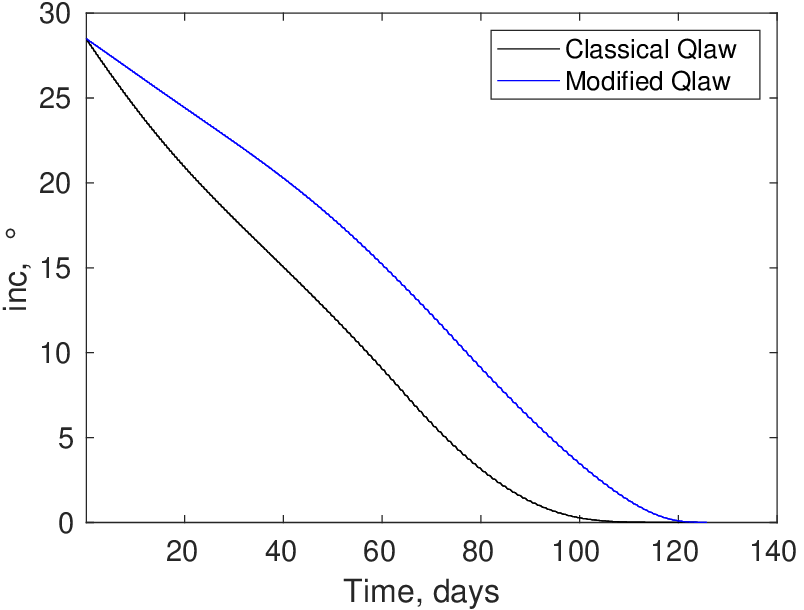}
    \subcaption{Inclination}\label{fig:inc_C}
  \end{minipage}
  \caption{GTO to GEO Transfer:: Trajectory trace comparison between Classical and Modified Qlaw}
  \label{fig:gto2geo1}
\end{figure}

\begin{figure}[hbt!]
  \centering
  \begin{minipage}{.23\textwidth}
    \centering
    \includegraphics[width=0.9\textwidth]{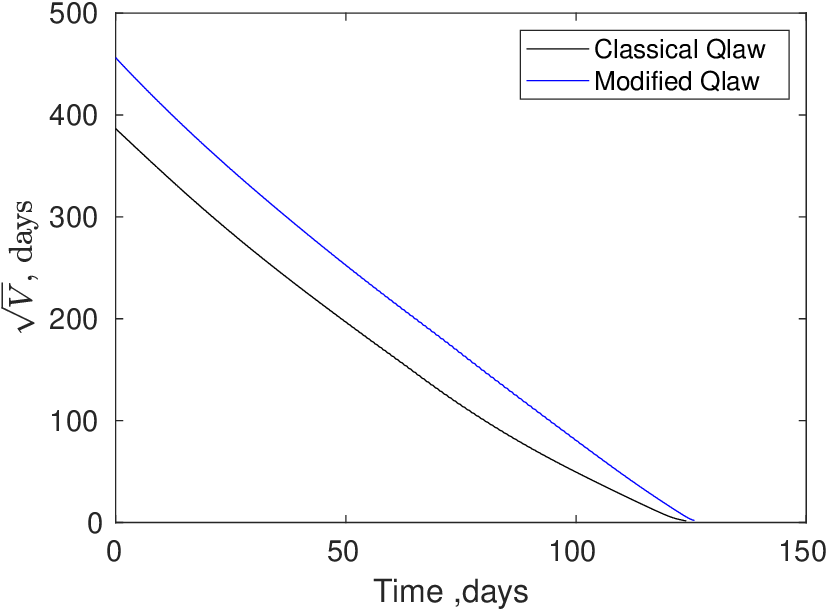}
    \subcaption{V}\label{fig:v_C}
  \end{minipage}
  \begin{minipage}{.23\textwidth}
    \centering
    \includegraphics[width=0.9\textwidth]{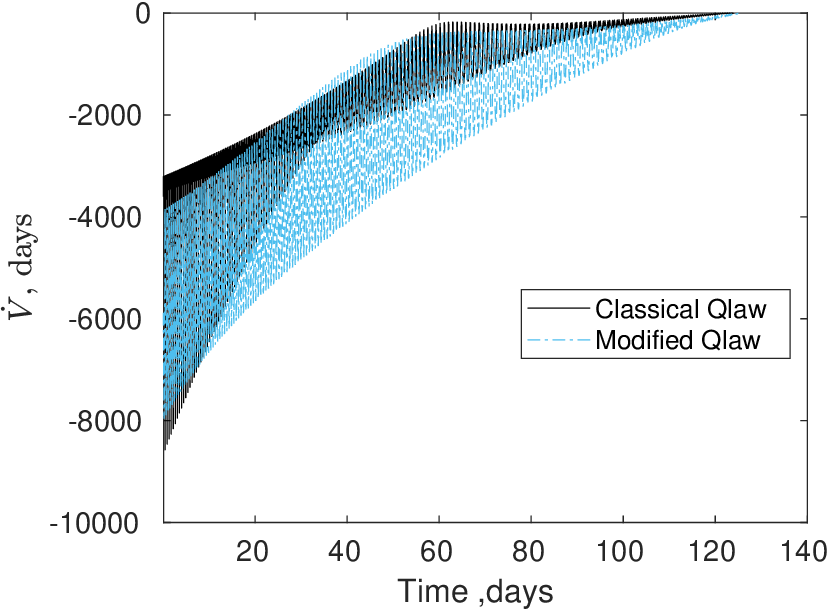}
    \subcaption{$\dot{V}$}\label{fig:vdot_C}
  \end{minipage}

  \caption{GTO to GEO Transfer:: Evolution of V and $\dot{V}$}
  \label{fig:gto2geo2}
\end{figure}

\begin{figure}[ht!]
  \centering
    \includegraphics[width=0.3\textwidth]{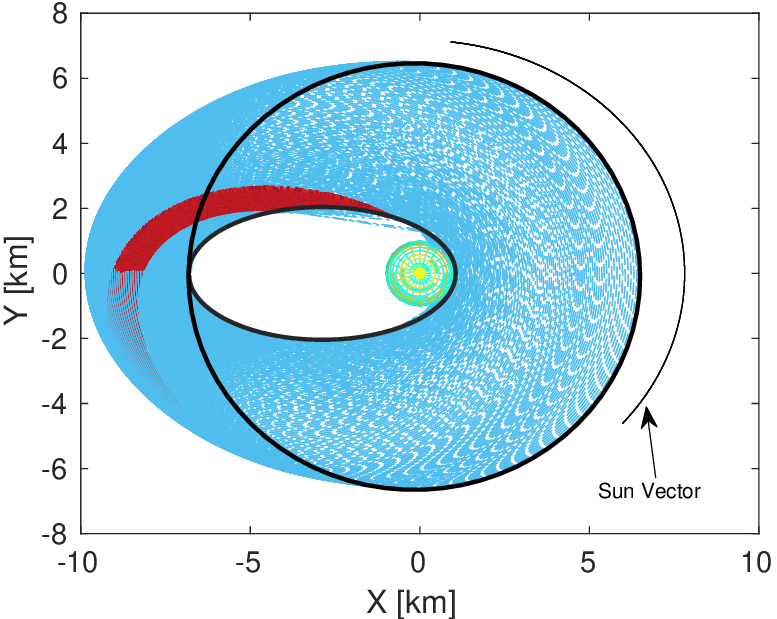} 
    \caption{GTO-GEO Transfer: Modified Q-law Trajectory Projected in Equatorial plane (Normalized by Earth's Radius)}
    \label{gto_geo_xy_orb}
\end{figure}

Finally, simulation results for the GTO-to-GEO transfer, representative of communication satellite missions, are presented. Figures~\ref{fig:v_C}–\ref{fig:vdot_C} illustrate the evolution of the Lyapunov function and its derivative for both control laws, while Fig.~\ref{fig:a_C}-\ref{fig:inc_C} shows the corresponding evolution of the orbital elements. Note that penalty function is not included in this transfer for classical Q-law for one to one comparision. It is evident that the modified Q-law achieves performance comparable to the classical Q-law without any degradation in performance metrics, while simultaneously ensuring closed-loop stability. Table \ref{tab:gto-geo-tf} shows the comparision of time-optimal results with existing literature\cite{shannon2020q}, \cite{graham2016minimum}, \cite{leomanni2021optimal}. Fig.~\ref{gto_geo_xy_orb} shows the projected trajectory for the complete transfer. The dark red patch shows the eclipse arc as sun moves with respect to spacecraft frame.
  
\begin{table}[hbt!]
\caption{\label{tab:gto-geo-tf} Time Optimal Results for GTO-GEO Transfer}
\centering
\begin{tabular}{lcc}
\hline
Method  & transfer time, days \\\hline
Minimum time Trajectory Optimization & 119-122  \\
Optimized Modified Q-law & 124.7   \\
\hline
\end{tabular}
\end{table}

\section{Conclusion and future work}  \label{sec:conc}
This paper presents a novel Lyapunov-based closed-loop control law for low-thrust, multiple-revolution orbit transfers that ensures closed-loop stability while remaining suitable for real-time onboard implementation. Simulation results demonstrate that the modified Q-law achieves performance comparable to the classical Q-law while guaranteeing stability. The present work focuses on the combined correction of semi-major axis, eccentricity, and inclination. However, for low Earth orbit maneuvers, correction of the right ascension of the ascending node (RAAN) is also crucial. To enable fully autonomous orbit maneuvers, future work will extend the proposed approach to include RAAN and argument of perigee corrections, as well as incorporate control parameters optimization to develop an adaptive control law. 

\section*{Acknowledgments}
We convey our sincere gratitude to the Director, U R Rao Satellite Center (URSC) for encouraging and supporting this research. We wish to gratefully acknowledge the excellent review committee at URSC for reviewing this work and providing valuable feedback. This research work was carried out as a part of Guidance Navigation and Control development for the All Electric Propulsion System Mission.
\bibliography{sample}

\end{document}